# Journalists' Perceptions of Artificial Intelligence and Disinformation Risks

Urko Peña-Alonso*, Simón Peña-Fernández[2] and Koldobika Meso-Ayerdi[3]

1. University of the Basque Country (UPV/EHU) (ROR: 000xsnr85)
   urko.pena@ehu.eus@ehu.eus / https://orcid.org/0000-0002-9214-5906
2. University of the Basque Country (UPV/EHU) (ROR: 000xsnr85)
   simon.pena@ehu.eus / https://orcid.org/0000-0003-2080-3241
3. University of the Basque Country (UPV/EHU) (ROR: 000xsnr85)
   koldo.meso@ehu.eus / https://orcid.org/0000-0002-0400-133X

* Corresponding author

**Abstract**: This study examines journalists' perceptions of the impact of artificial intelligence (AI) on disinformation, a growing concern in journalism due to the rapid expansion of generative AI and its influence on news production and media organizations. Using a quantitative approach, a structured survey was administered to 504 journalists in the Basque Country, identified through official media directories and with the support of the Basque Association of Journalists. This survey, conducted online and via telephone between May and June 2024, included questions on sociodemographic and professional variables, as well as attitudes toward AI's impact on journalism. The results indicate that a large majority of journalists (89.88%) believe AI will considerably or significantly increase the risks of disinformation, and this perception is consistent across genders and media types, but more pronounced among those with greater professional experience. Statistical analyses reveal a significant association between years of experience and perceived risk, and between AI use and risk perception. The main risks identified are the difficulty in detecting false content and deepfakes, and the risk of obtaining inaccurate or erroneous data. Co-occurrence analysis shows that these risks are often perceived as interconnected. These findings highlight the complex and multifaceted concerns of journalists regarding AI's role in the information ecosystem.

## 1. Introduction

In the past decade, disinformation has emerged as one of the main concerns in the field of journalism (Cea & Palomo, 2021). This phenomenon has gained a relevance comparable to the exponential increase in the strategic dimension of journalistic practice, driven by the expansion of artificial intelligence (Palomo et al., 2023; Verma, 2024; Kevin-Alerechi et al., 2025).

Thus, the advances in artificial intelligence (AI) have surprised us in recent years due to their speed and their impact on almost every field. Generative AI, which marked a turning point in the development of this technology, is particularly relevant in content creation spaces (Gonzalo, 2024; Newman et al., 2024). In the field of journalism, it is increasingly evident that AI is beginning to make inroads transversally in the news production process, as well as in the structure and functioning of media organizations (Túñez López et al., 2021; Ufarte Ruiz et al., 2023; Londoño-Proaño & Buele, 2025). The predictions of those who foresaw that AI would ultimately determine the future development of media transformation seem to be coming true (Zheng et al., 2018; Cui, 2025).

Peña-Fernández et al. (2023) state that, although AI brings benefits to journalism, its adoption entails ethical challenges that must be proactively addressed. Its emergence has rekindled the debate regarding the ambivalent nature of technological developments (Brennen et al., 2018), that is, the need to assess the balance between their potential benefits and the new insecurities they introduce (Vrabič Dežman, 2024; Wei et al., 2024; Schaetz et al., 2025).

It is evident that the use of artificial intelligence and its application to journalism reveal significant opportunities for improvement in the efficiency of news production and distribution, but also the threats posed by its use to undermine quality journalism (Mateos Abarca & Gamonal Arroyo, 2024) or to disseminate disinformation on a large scale (Astobiza, 2024). These challenges are compounded by concerns regarding the veracity of content, ethics and transparency in data usage, intrusion into private life, and its contribution to reinforcing existing social divides (Hansen et al., 2017; Brundage et al., 2018; Pihlajarinne & Alén-Savikko, 2022).

Media professionals themselves are not immune to this debate. On the one hand, AI can assist journalists by freeing them from the most routine and low value-added tasks (Wu et al., 2019; Cools & Diakopoulos, 2024; Cools & de Vreese, 2025; García de Torres et al., 2025), which could help strengthen the more cognitive aspects of the profession, such as creativity, critical thinking, curiosity, or skepticism (Thurman et al., 2017; Guzman & Lewis, 2020; Moravec et al., 2024). For media organizations, AI may serve as the key to accessing business models that improve their economic viability (Lindén & Tuulonen, 2019; Kotenidis & Veglis, 2021; Oh & Jung, 2025) or help them reduce costs (Wilczek et al., 2024).

However, the main reservations regarding the implementation of AI come from the workers themselves, who see it more as a threat than as a complement (Mondría Terol, 2023; van Dalen, 2024). In fact, some studies conclude that media managers adopt automation as a strategy aimed primarily at reducing costs and improving productivity, resulting in a reduction in the number of employees (Kim & Kim, 2017; Rick & Hanitzsch, 2023; Thäsler-Kordonouri & Barling, 2023) or an increase in job insecurity (López Jiménez & Ouariachi, 2020; Ananny & Karr, 2025).

In any case, the aim is to make the journalist's work more efficient (Papadimitriou, 2016; Dodds et al., 2025), which makes its implementation in the media essential (Autor, 2015; Rostamian & Moradi, 2024). At the same time, the adoption of AI demands new professional profiles (Sixto García et al., 2021; Møller et al., 2024), which require not only learning to use certain tools (Murcia Verdú & Lara Ramos, 2024), but also focusing their training on three central pillars: foundational knowledge, technical skills, and ethical competencies (Lopezosa et al., 2023). In this regard, various authors argue that AI literacy will be a fundamental requirement for future journalists (Graefe, 2016; Sonni et al., 2024; Sonni, 2025).

The integration of AI in journalism is not without ethical challenges. Beyond the potential benefits attributed to it, one of the most debated dilemmas is the inherent bias in AI algorithms (Binns, 2018; Deuze & Beckett, 2022), compounded by the risk of large-scale dissemination of disinformation (Kertysova, 2018). Additionally, there is the threat faced by news consumers in discerning the source and veracity of information, which can lead to trust issues in the media (Montal & Reich, 2017; Wu, 2024; Morosoli et al., 2025).

AI can facilitate certain specific activities within the news production process (Canavilhas, 2022), among which the fight against disinformation stands out (Manfredi Sánchez & Ufarte Ruiz, 2020; Saeidnia et al., 2025), a matter of particular concern in democratic countries (Rodríguez Martelo et al., 2023). In fact, the advent of AI has increased the possibilities for combating disinformation (Moreno Espinosa et al., 2024) and can help distinguish between truthful information and the distortion of reality (Flores Vivar, 2019; Lange & Lechterman, 2021; Santos, 2023).

Ballesteros and del Olmo (2024) state that AI has the capacity to analyze large amounts of information quickly and efficiently, identifying patterns and anomalies that can reveal the veracity of a news item.

In this regard, some journalistic organizations have expressed their distrust regarding the evolution of the sector, due to the lack of capacity to combat disinformation and unverified information (Subiela Hernández et al., 2023). To address this concern, fact-checkers have emerged as contextual agents to shape a new ecosystem that fights against disinformation (Pilo García et al., 2024). Indeed, some advocate for the incorporation of AI into content-verification processes (Tejedor & Sancho Ligorred, 2023). Thus, the potential of AI for information verification—a key process in communication—is recognized (Cuartielles et al., 2023), and it has been demonstrated that the use of AI can be vital for counteracting the disruptive effects of disinformation (Rubin, 2022), as it reduces detection time and increases the capacity to respond to it (Cuartielles et al., 2024).

Unfortunately, these same AI tools can also prove effective in amplifying the reach and impact of disinformation by enabling the generation of false content. The misuse of AI tools represents a significant challenge in the information age (La Rosa & Luján, 2024), as it is possible, at a relatively low cost, to use AI to distribute false information worldwide. This possibility is seen as a threat by the public and by a large part of information professionals, since messages produced using artificial intelligence (AI) techniques can be falsely attributed to non-existent information sources, incorporate false or inaccurate content, and be used for partisan or malicious purposes (Peña-Fernández et al., 2023).

The problem of disinformation, understood as that which serves a specific purpose and includes an intentional component, poses a greater risk when situated within the context of a hyperconnected society, where any user of digital environments can simultaneously be both a victim and a conscious or unconscious source of disinformation (Benaissa Pedriza, 2024, pp. 106–107). The malicious use of these models to generate false and credible content automatically, massively, and at no cost constitutes one of their main risks and is the most immediate argument for those calling for urgent regulation (Sanguinetti, 2023, p. 13).

Moreover, AI not only has the capacity to create convincing and realistic content, but it also raises serious ethical challenges in multiple areas, especially regarding privacy, data protection, and copyright, among others (García, 2024).

In fact, the emergence of AI in journalism raises questions about the ethics, quality, and transparency of automated information processes. It is crucial that journalism professionals and media organizations understand these challenges and work to address them effectively (Esteban Regules & Calle Mendoza, 2024, p. 251).

In this context, this article aims to analyze Basque journalists' perceptions of the impact of artificial intelligence on the phenomenon of disinformation. To address this main objective, this research poses the following questions:
RQ1. How is AI going to impact information disorders (disinformation, false content, and deepfakes)?
RQ2. Are there significant differences among journalists based on variables such as gender, years of professional experience, the type of media in which they work, and their position within the organization?
RQ3. How do Basque journalists perceive risks in their professional use of AI? At which levels—individual and associated—are these risks perceived?

## 2. Materials and methods

This study adopts a quantitative approach, based on a structured survey administered to journalists working in the Basque Country. The sample consists of 504 journalists, identified through the Open Communication Guide of the Basque Government (Basque Government, n.d.), which lists the active media outlets in the region, and with the collaboration of the Basque Association of Journalists (Basque Association of Journalists—Basque College of Journalists, n.d.). The distribution of the survey also benefited from the support of this association. Of the participants, 276 are men (55.1%), 226 are women (44.5%), and 2 individuals identified with other genders (0.4%). According to available data (Basque Government, 2022; Pérez et al., 2023), it is estimated that approximately 5000 individuals are employed in the media sector in Euskadi. Therefore, for a 95% confidence level, the survey's margin of error is ±4.15% (Sarrionandia et al., 2025).

Data acquisition was carried out between May and June 2024 through anonymous surveys, without recording any personally identifiable information or sensitive data. At no point were respondents asked to provide their name, address, email, telephone number, or any other information that could allow for their direct or indirect identification.

Participation in the survey was entirely voluntary and based on informed consent, which was provided at the beginning of the questionnaire. This introductory section clearly explained the purpose of the research, the confidentiality of responses, and the respondent's right to decline to complete the form if they wished.

As the study involves no physical or psychological risks, does not process specially protected data, and complies with the provisions of the General Data Protection Regulation (GDPR) and current national legislation, it is considered that formal ethics committee approval is not required, in line with standard guidelines for low-risk research in academic settings.

The questionnaire was administered primarily through an online platform, which maximized geographic coverage and participant accessibility, and was complemented by telephone support to facilitate participation.

The survey was developed ad hoc for the present study. Prior to its administration, the questionnaire underwent a validation process, which included a pilot test with a small group of journalists to ensure the clarity, relevance, and comprehensibility of the items.

The questionnaire included closed and multiple-choice questions, organized into two main sections: (a) sociodemographic and professional variables (gender, years of experience, type of media outlet, province of workplace, main activity in the media, and level of responsibility); and (b) perceptions and attitudes regarding the impact of AI on journalism. Among the specific issues analyzed are perceptions of the increase in the risks of information disorders (disinformation, false content, and deepfakes), the rise of biases and digital divides (gender, social class, etc.), and the identification of the main risks associated with the use of AI in the journalistic field (allowing for a maximum of two options to be selected).

Only fully completed questionnaires were included in the analysis. Partially completed surveys and those missing informed consent were excluded on that basis.

The heatmaps presented in this study were generated using ChatGPT 4.1. All statistical analyses, including correspondence analysis (CA), were performed using R (version 4.4.0) within the RStudio integrated development environment (IDE) (version 1.1.456).

Figure 1. Percentage distribution of perceived disinformation risk by media type (X-axis: perception attributes; Y-axis: media type).

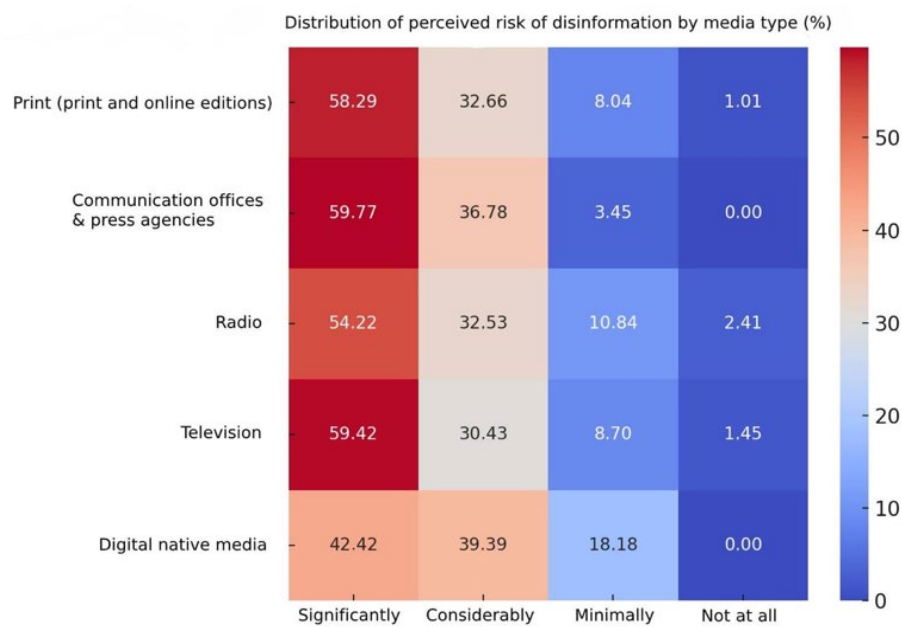

Source: Author's own work

## 3. Results

*3.1. Journalists' Perceptions of Disinformation Associated with AI*

The majority of professionals perceive that artificial intelligence will significantly increase the risks of disinformation. Specifically, 89.88% of the responses fall into the categories "considerably" (n = 171) and "significantly" (n = 282), while only 10.12% consider that the impact to fall under the categories of "minimally" (n = 46) or "not at all" (n = 5). These results suggest that journalists view artificial intelligence as a significant challenge to the integrity of the information ecosystem.

When analyzing the distribution of responses according to the type of media in which they work (Figure 1), a trend similar to that of the overall sample is observed. However, among journalists working in digital-native media, although the negative perception remains predominant (81.81%, n = 54), there is a higher percentage of responses that minimize the impact ("minimally": 18.18%) compared to the average for other categories, such as print, radio, or television (91.03%). In contrast, professionals working in communication offices and press agencies are the most likely to consider that AI will have a high negative impact, with 96.95% (n = 84) in the highest-impact categories.

The perception of the risk of AI-driven disinformation is very similar across genders, with 90.22% (n = 249) for men and 89.82% (n = 196) for women in terms of relative frequency for the "significantly" and "considerably" options (Figure 2). The "other" category includes only two cases, so its representativeness is very limited.

Figure 2. Percentage distribution of perceived disinformation risk by gender (X-axis: gender; Y-axis: perception attributes).

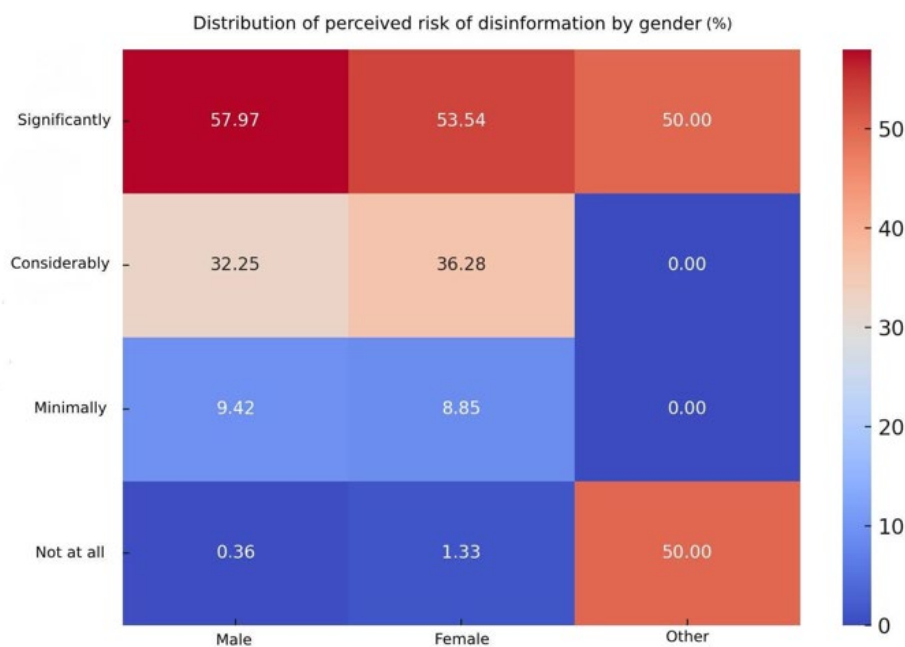

Source: Author's own work.

The results of Fisher's exact test, with a p-value of 0.05419, are not conclusive, as they are slightly above the reference value of 0.05. After excluding the "other" category and grouping the "minimally" and "not at all" categories to ensure a sufficiently large sample size per category (typically above 30 cases) for the Chi-square test, the p-value increases to 0.5916. The effect size, measured by Cramér's V, is 0.046, which is considered very small (Cohen, 1988). Adjusting for multiple comparisons using the Holm and Benjamini–Hochberg methods did not alter this conclusion (adjusted p-values: 0.5916). Therefore, we can conclude that there is not sufficient statistical evidence of a relationship between these variables.

With regard to the relationship between information disorders and years of experience in journalism (Figure 3), the sum of the percentages for the "significantly" and "considerably" options for each of the groups studied suggests a trend with some nuances: as years of experience increase, so does the perception of AI-related disinformation risk (more than 20 years, 92.76%; 10 to 20 years, 94.44%; 5 to 10 years, 87.32%; and five years or less, 80.77%). Journalists with five years or less of professional experience report the lowest percentage in risk perception; however, this group still represents a clear majority overall.

Figure 3. Percentage distribution of perceived disinformation risk by years of experience (X-axis: years of experience; Y-axis: perception attributes).

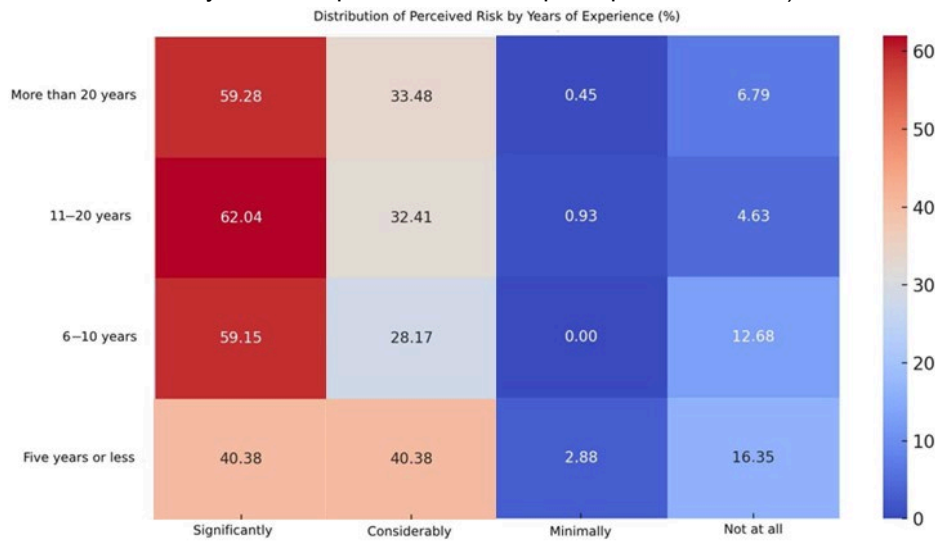

Source: Author's own work.

In this case, the p-value (0.001937) from Fisher's exact test clearly indicates the existence of statistically significant evidence; these data support the conclusion that there is an association between the variables studied. The effect size (Cramér's V = 0.144) is small to medium, and the result remains significant after adjusting for multiple comparisons (adjusted p-value: 0.0058, Holm; 0.0058, Benjamini–Hochberg).

The correspondence analysis visualization (Figure 4) corroborates the previously noted trend. As professional experience in the field increases, journalists are more likely to perceive a greater influence of AI on disinformation. Nevertheless, all journalists, regardless of years of experience, attribute some impact to AI, as can be inferred from the distance between the "not at all" category (red triangle) and its counterparts representing years of experience (blue dots). Regarding the correspondence analysis, the first dimension explained 85.7% of the total inertia, while the second dimension accounted for 13.3%. Together, the first two dimensions explained 99.0% of the total variance, indicating that the two-dimensional solution provides a good summary of the association structure in the data.

With regard to how the use of artificial intelligence in the workplace influences the perception of disinformation risks, Fisher's exact test yielded a p-value of 0.0056, indicating a statistically significant association between the variables. The effect size, measured by Cramér's V, was 0.120, which is considered small to medium according to Cohen's (1988) criteria. This result remained significant after adjusting for multiple comparisons using both the Holm (adjusted p = 0.0112) and Benjamini–Hochberg (ad-justed p = 0.0084) methods.

Figure 4. Correspondence analysis of the association between journalistic experience and perceived influence of artificial intelligence on disinformation.

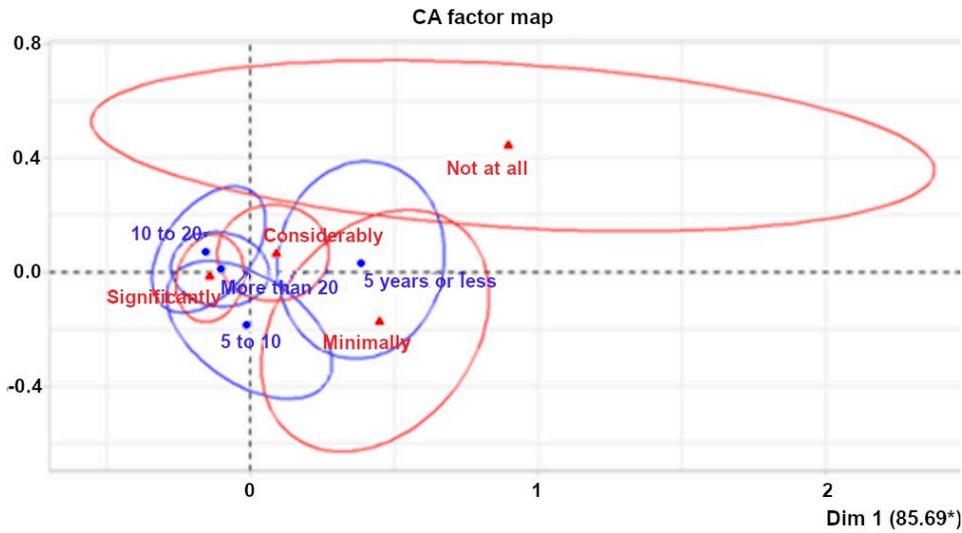

Source: Author's own work.

As shown in the correspondence analysis (Figure 5), there is an inverse relationship between journalists who do not use artificial intelligence at all and their perception of its influence on disinformation, as they tend to choose the "significantly" category. In contrast, information professionals who make intensive use of this technology have a more moderate perception of its impact. The correspondence analysis revealed that Dimension 1 accounted for 70.3% of the total inertia, and Dimension 2 contributed an additional 27.6%. Combined, these two dimensions captured 97.9% of the overall variance, demonstrating that the relationship between the variables is effectively represented within a two-dimensional space.

Figure 5. Correspondence analysis of artificial intelligence usage in the workplace and journalists' perceptions of disinformation risk.

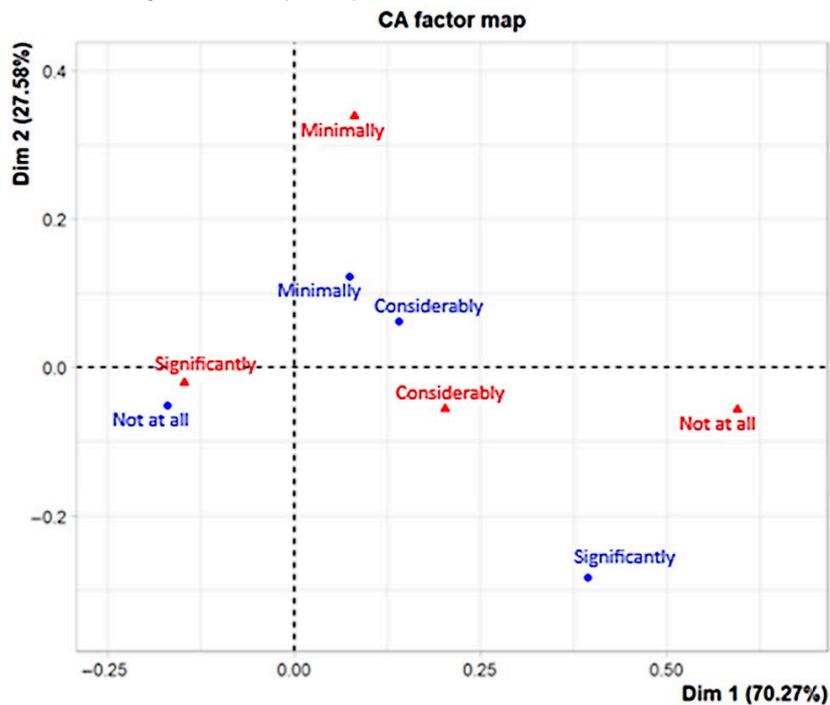

Source: Author's own work.

*3.2. Perceived Risks of AI Use Among Basque Journalists*

Regarding the specific risks associated with the direct use of artificial intelligence, Basque journalists mainly highlight the difficulty in identifying false content and deepfakes, as well as the risk of obtaining inaccurate or erroneous data. Difficulty detecting false content and deepfakes is the most frequently cited risk, accounting for 37.9% of responses (n = 382), followed by concerns about content or data inaccuracy, mentioned by 33.33% (n = 336). Other risks include becoming a victim of criminal uses, such as scams or skimming (11.61%, n = 117), and biases stemming from data origin (gender, social class, etc.) (10.81%, n = 109). Only 6.35% (n = 64) (Table 1) report no additional risks. It is important to note that the total number of responses (1008) exceeds the number of respondents (504), as each participant was allowed to select up to two risks. This design permitted respondents to indicate more than one perceived risk, resulting in a higher overall response count.

Table 1. Main risks identified by journalists in their use of AI.

| **Risk Category** | **Count** | **Percentage** |
|---|---|---|
| Difficulty identifying false content and deepfakes | 382 | 37.9% |
| Obtaining inaccurate or erroneous content/data | 336 | 33.33% |
| Being a victim of criminal uses (scams and skimming) | 117 | 11.61% |
| Biases due to data origin (gender, social class, etc.) | 109 | 10.81% |
| None other | 64 | 6.35% |

Note: Respondents could select up to two risks.
Source: Author's own work.

To examine how the different risks are associated in respondents' minds, we built a 4 × 4 co-occurrence matrix that combines the two options each journalist could select. Table 2 reports the ten distinct risk pairs and their relative frequencies; the three most frequent pairs account for more than three quarters of all co-occurrences.

Table 2. Co-occurrence of perceived risk pairs in journalists' use of AI.

| **Risk Pair** | **Count** | **Percentage** |
|---|---|---|
| Difficulty identifying false content and deepfakes paired with obtaining inaccurate or erroneous content/data | 253 | 50.5% |
| Difficulty identifying false content and deepfakes paired with being a victim of criminal uses (scams, skimming) | 69 | 13.77% |
| Obtaining inaccurate or erroneous content/data paired with biases due to data origin (gender, social class, etc.) | 58 | 11.58% |
| Difficulty identifying false content and deepfakes paired with biases due to data origin (gender, social class, etc.) | 40 | 7.99% |
| Obtaining inaccurate or erroneous content/data paired with being a victim of criminal uses (scams, skimming) | 17 | 3.99% |
| Difficulty identifying false content and deepfakes paired with none other | 20 | 3.99% |
| Being a victim of criminal uses (scams, skimming) paired with none other | 25 | 4.99% |
| Obtaining inaccurate or erroneous content/data paired with none other | 8 | 1.6% |
| Biases due to data origin (gender, soc. class, etc.) paired with none other | 5 | 1% |
| Biases due to data origin (gender, social class, etc.) paired with being a victim of criminal uses (scams, skimming) | 6 | 1.2% |

Note: Because each respondent could select up to two risks,
the co-occurrence counts are based on 504 respondents.
Source: Author's own work.

Among the journalists who selected two risks, 50.5% (n = 253) combined concerns about difficulty identifying false content and deepfakes with the risk of obtaining inaccurate or erroneous content or data. This finding suggests that, for a significant portion of respondents, the challenges of detecting false content, such as deepfakes, manipulated images, or fabricated news, are closely linked to the risk of receiving inaccurate or erroneous information generated by AI systems. In other words, journalists do not view these issues in isolation; rather, they perceive them as part of a broader, interconnected threat to the reliability of information.

The second most common pairing, selected by 13.8% (n = 69) of respondents, links the difficulty of identifying false content and deepfakes with the risk of being a victim of criminal uses, such as scams or skimming.

The third most frequent pairing, identified by 11.6% (n = 58) of respondents, combines concerns about obtaining inaccurate or erroneous content or data, with worries about biases due to data origin, such as gender, social class, or other demographic factors. This co-occurrence indicates that, for a segment of journalists, the issue of factual correctness is closely intertwined with the question of fairness in AI-generated information. A related finding is that 7.99% (n = 40) of respondents also paired difficulty identifying false content and deepfakes with concerns about data biases.

The remaining pairs have markedly lower frequencies. Two involve "none other", suggesting a subset of respondents who perceive only a single additional risk beyond their primary choice. Finally, co-occurrences between biases and criminal uses (1.2%) or between inaccuracy and criminal uses (3.99%) appear only sporadically.

## 4. Discussion

In line with the findings of previous studies, the results of this research show a broad consensus among journalists regarding the perceived risk that the implementation of artificial intelligence may pose in amplifying disinformation (RQ1). This view gives rise to predominantly cautious attitudes toward the application of this technology in journalism (Peña-Fernández et al., 2023), and it highlights the need for an ethical approach, as well as oversight of the content produced in this way (Noain-Sánchez, 2022).

In contrast to discourses that emphasize the potential of applying this technology for journalistic purposes (De-Lara-González et al., 2022; Mondría Terol, 2023), concern over its potential malicious use is widespread among media professionals. This aligns with the perception that it could eventually become a "weapon of mass disinformation", capable of eroding trust in journalism, weakening democracy, and manipulating public opinion on a large scale (Kertysova, 2018; Goldstein et al., 2023; D'Andrea et al., 2025).

Within the broad professional consensus on this issue, certain nuances can be identified. Thus, although the analysis reveals a notable gender balance among the journalists surveyed (RQ2), the perceived risk is somewhat lower among those working in native digital media and among younger practitioners. This suggests that more modern outlets and digital-native journalists adopt a slightly less pessimistic stance than their older counterparts and legacy media organizations. In this regard, it is important to note that when cross-analyzing the variables of years of experience and type of media, no evidence was found to suggest that journalists working in digital-native outlets have less professional experience than their counterparts in print, radio, or television. This suggests that the observed differences in risk perception are not merely a reflection of varying levels of experience across media types.

This trend, supported by statistically significant evidence, may be explained by greater exposure to the evolution of technological tools and a deeper understanding of the mechanisms through which disinformation can be generated and disseminated, or by generational differences in attitudes toward technological change (Manfredi Sánchez & Ufarte Ruiz, 2020). It can be understood as a form of resistance to change, but also as a well-founded warning from those who are most deeply familiar with the standards of the profession. In a context where media professionals are generally the most cautious and reluctant to embrace AI (Kim & Kim, 2018), this more open attitude toward its impact on disinformation among younger professionals and those in digital-native environments may signal a future shift in the perception of this threat, both within the profession and among audiences (Ross Arguedas, 2024).

This possible trend is further supported by the finding that the more frequently AI is used by journalism professionals, the lower their perception of risk tends to be. Journalists who do not use artificial intelligence in their daily work tend to perceive its impact on disinformation as particularly significant, whereas those who make intensive use of AI tools generally adopt a more moderate view regarding the risks associated with this technology. This pattern suggests that direct experience and familiarity with AI may foster a more nuanced understanding of both its potential and its limitations (Gutiérrez-Caneda et al., 2024).

The analysis of perceived risks associated with the use of artificial intelligence among journalists (RQ3) also reveals a clear hierarchy of concerns, in which worries extend beyond technocentric views to focus on the potential impact of this technology on both the public and the journalistic profession itself (Peña-Fernández et al., 2023). The difficulty in identifying false content and deepfakes emerges as the most prominent risk (Lundberg & Mozelius, 2025), closely followed by concerns about obtaining inaccurate or erroneous data (Nguyen, 2023). Other notable risks include the potential for criminal uses—such as scams or skimming—and the presence of biases arising from the origin of the data, particularly those related to gender or social class (Leiser, 2022).

In the same vein, the analysis of risk pairings underscores that, for a significant portion of journalists, the challenges of detecting manipulated or fabricated information are not perceived in isolation. Rather, they are seen as intrinsically linked to the broader issue of the reliability and accuracy of AI-generated content, as well as to the vulnerability of both media professionals and audiences in the context of rapid technological transformation.

This research contributes to a deeper understanding of how journalistic cultures respond to perceived emerging technological threats. By uncovering nuanced attitudes among professionals with diverse backgrounds and roles, this study reveals that, despite the strong consensus among journalists regarding the risks posed by AI in amplifying disinformation, perceptions are also shaped by generational and technological factors. These underlying differences may offer valuable insight into the future evolution of professional attitudes, suggesting that greater exposure to and familiarity with AI could gradually lead to more balanced and informed assessments of its impact within journalism. Therefore, promoting AI literacy and targeted training among media professionals could be key to fostering a critical—but not reactive—adoption of this technology.

This study presents several limitations that should be acknowledged. The geographic focus on the Basque Country—characterized by a distinctive media landscape and sociolinguistic context—means that the findings may reflect specific regional dynamics and should be interpreted with caution when applied to broader contexts. In addition, the cross-sectional design offers a snapshot of journalists' perceptions at a particular moment in time, which could evolve as artificial intelligence becomes increasingly embedded in journalistic practices. Finally, as participation in the survey was voluntary, there is the possibility of a slight self-selection effect, whereby individuals with stronger views on AI and disinformation may have been more motivated to respond.

Looking ahead, several directions for future research could help to deepen and broaden the insights offered by this study. A longitudinal design may be useful to explore how perceptions of risk shift over time, particularly as journalists gain greater exposure to and familiarity with AI technologies. In addition, extending the study to other European regions—especially those with peripheral or multilingual media systems—could provide valuable comparative perspectives and help contextualize the findings. Finally, examining the role of training and algorithmic literacy in shaping risk perceptions may offer further nuance, particularly in understanding whether targeted technical or ethical education contributes to more informed and balanced views among media professionals.

## References

Ananny, M., & Karr, J. (2025). How media unions stabilize technological hype: Tracing organized journalism's discursive constructions of generative artificial intelligence. Digital Journalism, 1–21. https://doi.org/10.1080/21670811.2025.2454516.

Astobiza, A. M. (2024). Deepfakes, desinformación, discursos de odio y democracia en la era de la Inteligencia Artificial. Cuadernos del Audiovisual/CAA, 12, 177–190. https://doi.org/10.62269/cavcaa.20.

Autor, D. H. (2015). Why are there still so many jobs? The history and future of workplace automation. Journal of Economic Perspectives, 29(3), 3–30. https://doi.org/10.1257/jep.29.3.3.

Ballesteros, L., & del Olmo, F. (2024). Vídeos falsos y desinformación ante la IA: El deepfake como vehículo de la posverdad. Revista de Ciencias de la Comunicación e Información, 29, 1–14. https://doi.org/10.35742/rcci.2024.29.e294.

Basque Association of Journalists—Basque College of Journalists. (n.d.). Kazetariak. Available online: https://kazetariak.eus/ (accessed on 22 March 2025).

Basque Government. (2022). Censo del mercado de trabajo. Specific Statistical Body of the Department of Labor and Employment. Available online: https://www.eustat.eus/elementos/ele0021400/censo-del-mercado-de-trabajo-oferta/inf0021420_c.pdf (accessed on 22 March 2024).

Basque Government. (n.d.). Open communication guide. Available online: https://gida.irekia.euskadi.eus (accessed on).

Benaissa Pedriza, S. (2024). Activistas, «influencers» y usuarios de redes sociales como fuente de desinformación: Una tipología operativa de nuevos líderes de opinión en los entornos digitales. Enrahonar, 73, 105–129. https://doi.org/10.5565/rev/enrahonar.1570.

Binns, R. (2018, February 23–24). Fairness in machine learning: Lessons from political philosophy. 2018 Conference on Fairness, Accountability, and Transparency (pp. 149–159), New York, NY, USA. https://doi.org/10.1145/3287560.3287583.

Brennen, J. S., Howard, P. N., & Nielsen, R. K. (2018). An industry-led debate: How UK media cover artificial intelligence. Reuters Institute for the Study of Journalism.

Brundage, M., Avin, S., Clark, J., Toner, H., Eckersley, P., Garfinkel, B., Dafoe, A., Scharre, P., Zeitzoff, T., Filar, B., Anderson, H., Roff, H., Allen, G. C., Steinhardt, J., Flynn, C., Ó hÉigeartaigh, S., Beard, S., Belfield, H., Farquhar, S., Lyle, C., Crootof, R., Evans, O., Page, M., Bryson, J., Yampolskiy, R., & Amodei, D.(2018). The malicious use of artificial intelligence: Forecasting, prevention, and mitigation. Available online: https://tinyurl.com/3ayc5tnw (accessed on).

Canavilhas, J. (2022). Inteligencia artificial aplicada al periodismo: Traducción automática y recomendación de contenidos en el proyecto "A European Perspective" (UER). Revista Latina de Comunicación Social, 80, 1–13. https://doi.org/10.4185/RLCS-2022-1534.

Cea, N., & Palomo, B. (2021). Disinformation matters. Analyzing the academic production. In G. López-García, D. Palau-Sampio, B. Palomo, E. Campos-Domínguez, & P. Masip (Eds.), Politics of disinformation: The influence of fake news on public sphere. John Wiley & Sons.

Cohen, J. (1988). Statistical power analysis for the behavioral sciences (2nd ed.). Lawrence Erlbaum Associates.

Cools, H., & de Vreese, C. H. (2025). From automation to transformation with AI-tools: Exploring the professional norms and the perceptions of responsible AI in a news organization. Digital Journalism, 1–20. https://doi.org/10.1080/21670811.2025.2505982.

Cools, H., & Diakopoulos, N. (2024). Uses of generative AI in the newsroom: Mapping journalists' perceptions of perils and possibilities. Journalism Practice, 1–19. https://doi.org/10.1080/17512786.2024.2394558.

Cuartielles, R., Mauri Ríos, M., & Rodríguez Martínez, R. (2024). Transparencia en el uso de la IA en las plataformas de fact-checking en España y sus desafíos éticos. Communication & Society, 37(4), 257–271. https://doi.org/10.15581/003.37.4.257-271.

Cuartielles, R., Ramon Vegas, X., & Pont Sorribes, C. (2023). Retraining fact-checkers: The emergence of ChatGPT in information verification. Profesional de la información, 32(5), 1-17. https://doi.org/10.3145/epi.2023.sep.15.

Cui, J. (2025). Digital transformation in the media industry: The moderating role of human-AI interaction technologies. Media, Communication, and Technology, 1(1), 42–47. https://doi.org/10.30560/mct.v1n1p42.

D'Andrea, A., Fusacchia, G., & D'Ulizia, A. (2025). Linguistic insights, media mechanisms, and the role of AI in dissemination and impact of disinformation. Journal of Information, Communication and Ethics in Society.

De-Lara-González, A., García-Avilés, J. A., & Arias-Robles, F. (2022). Implantación de la inteligencia artificial en los medios españoles: Análisis de las percepciones de los profesionales. Textual & Visual Media, 1(15), 1–16. https://doi.org/10.56418/txt.15.2022.001.

Deuze, M., & Beckett, C. (2022). Imagination, algorithms, and news: Developing AI literacy for journalism. Digital Journalism, 10(10), 1913–1918. https://doi.org/10.1080/21670811.2022.2119152.

Dodds, T., Zamith, R., & Lewis, S. C. (2025). The AI turn in journalism: Disruption, adaptation, and democratic futures. Journalism. https://doi.org/10.1177/14648849251343518.

Esteban Regules, B., & Calle Mendoza, S. (2024). La transformación del periodismo: Ética y responsabilidad en el uso de la inteligencia artificial. In S. Mayorga Escala (Ed.), Tendencias de investigación en comunicación (pp. 251–266). Dykinson.

Flores Vivar, J. M. (2019). Inteligencia artificial y periodismo: Diluyendo el impacto de la desinformación y las noticias falsas a través de los bots. Doxa Comunicación, 29, 197–212. https://doi.org/10.31921/doxacom.n29a10.

García, A. M. (2024). ¿Qué desafíos éticos plantea la inteligencia artificial generativa? Ciencia Vital, 2(4), 1-2. https://doi.org/10.20983/cienciavital.2024.04.nde.01.

García de Torres, E., Ramos, G., Yezers' ka, L., Gonzales, M., Higuera, L., & Herrera, C. (2025). The use and ethical implications of artificial intelligence, collaboration, and participation in local Ibero-American newsrooms. Frontiers in Communication, 10, 1539844. https://doi.org/10.3389/fcomm.2025.1539844.

Goldstein, J. A., Sastry, G., Musser, M., DiResta, R., Gentzel, M., & Sedova, K. (2023). Generative language models and automated influence operations: Emerging threats and potential mitigations. arXiv, arXiv:2301.04246.

Gonzalo, M. (2024). Cuánto puede impactar la IA en la desinformación: El periodismo en la era de los 'deepfakes'. Cuadernos de periodistas: Revista de la Asociación de la Prensa de Madrid, 48, 69–77.

Graefe, A. (2016). Guide to automated journalism. Tow Center for Digital Journalism. Available online: https://academiccommons.columbia.edu/doi/10.7916/D8FJ2DWB (accessed on).

Gutiérrez-Caneda, B., Lindén, C. G., & Vázquez-Herrero, J. (2024). Ethics and journalistic challenges in the age of artificial intelligence: Talking with professionals and experts. Frontiers in Communication, 9, 1465178.

Guzman, A. L., & Lewis, S. C. (2020). Artificial intelligence and communication: A human–machine communication research agenda. New Media & Society, 22(1), 1–18. https://doi.org/10.1177/1461444819858691.

Hansen, M., Roca-Sales, M., Keegan, J., & King, G. (2017). Artificial intelligence: Practice and implications for journalism. Columbia Journalism School. https://doi.org/10.7916/D8X92PRD.

Kertysova, K. (2018). Artificial Intelligence and disinformation. Security and Human Rights, 29(1–4), 55–81. https://doi.org/10.1163/18750230-02901005.

Kevin-Alerechi, E., Abutu, I., Oladunni, O., Osanyinro, E., Ojumah, O., & Ogundele, E. (2025). AI and the newsroom: Transforming journalism with intelligent systems. Journal of Artificial Intelligence, Machine Learning and Data Science, 3(1), 1930–1937. https://doi.org/10.51219/JAIMLD/Elfredah-Kevin-Alerechi/426.

Kim, D., & Kim, S. (2017). Newspaper companies' determinants in adopting robot journalism. Technological Forecasting and Social Change, 117, 184–195. https://doi.org/10.1016/j.techfore.2016.12.002.

Kim, D., & Kim, S. (2018). Newspaper journalists' attitudes towards robot journalism. Telematics and informatics, 35, 340-357. https://doi.org/10.1016/j.tele.2017.12.009

Kotenidis, E., & Veglis, A. (2021). Algorithmic journalism—Current applications and future perspectives. Journalism and Media, 2(2), 244–257. https://doi.org/10.3390/journalmedia2020014.

Lange, B., & Lechterman, T. M. (2021, October 28–31). Combating disinformation with AI: Epistemic and ethical challenges. 2021 IEEE International Symposium on Technology and Society (ISTAS) (pp. 1–5), Waterloo, ON, Canada. https://doi.org/10.1109/ISTAS52410.2021.9629122.

La Rosa, A., & Luján, J. (2024). Del "periodismo de verdad" a las fake news en la era de la inteligencia artificial. Revista Científica de Comunicación Social, 6), 86–98. https://doi.org/10.71187/brc.v0i6.107.

Leiser, M. R. (2022). Bias, journalistic endeavours, and the risks of artificial intelligence. In T. Pihlajarinne, & A. Alén-Savikko (Eds.), Artificial intelligence and the media (pp. 8–32). Elgar. https://doi.org/10.4337/9781839109973.00007.

Lindén, C. G., & Tuulonen, H. (2019). News automation: The rewards, risks, and realities of 'machine journalism'. WAN-IFRA Report. Available online: https://tinyurl.com/yc24pmxc (accessed on).

Londoño-Proaño, C., & Buele, J. (2025). Can artificial intelligence replace journalists? A theoretical approach. Frontiers in Communication, 10, 1537146. https://doi.org/10.3389/fcomm.2025.1537146.

Lopezosa, C., Codina, Ll., Pont-Sorribes, C., & Vállez, M. (2023). Use of generative artificial intelligence in the training of journalists: Challenges, uses, and training proposal. Profesional de la Información, 32(4), e320408. https://doi.org/10.3145/epi.2023.jul.08.

Lundberg, E., & Mozelius, P. (2025). The potential effects of deepfakes on news media and entertainment. AI & Soc, 40, 2159–2170. https://doi.org/10.1007/s00146-024-02072-1.

López Jiménez, E., & Ouariachi, T. (2020). An exploration of the impact of artificial intelligence (AI) and automation for communication professionals. Journal of Information, Communication and Ethics in Society, 19(2), 249–267. https://doi.org/10.1108/JICES-03-2020-0034.

Manfredi Sánchez, J. L., & Ufarte Ruiz, M. J. (2020). Inteligencia artificial y periodismo: Una herramienta contra la desinformación. Revista CIDOB d'Afers Internacionals, 124, 49–72. http://doi.org/10.24241/rcai.2020.124.1.49.

Mateos Abarca, J. P., & Gamonal Arroyo, R. (2024). Metodologías de investigación y usos de la inteligencia artificial aplicada al periodismo. Comunicación y Métodos—Communication & Methods, 6(1), 90–107. https://doi.org/10.35951/v6i1.220.

Mondría Terol, T. (2023). Innovación mediática: Aplicaciones de la inteligencia artificial en el periodismo en España. Textual & Visual Media, 17(1), 41–60. https://doi.org/10.56418/txt.17.1.2023.3.

Montal, T., & Reich, Z. (2017). I, robot. You, journalist. Who is the author? Authorship, bylines and full disclosure in automated journalism. Digital Journalism, 5(7), 829–849. https://doi.org/10.1080/21670811.2016.1209083.

Moravec, V., Hynek, N., Skare, M., Gavurova, B., & Kubak, M. (2024). Human or machine? The perception of artificial intelligence in journalism, its socio-economic conditions, and technological developments toward the digital future. Technological Forecasting and Social Change, 200, 123162. https://doi.org/10.1016/j.techfore.2023.123162.

Moreno Espinosa, P., Abdulsalam Alsarayreh, R. A., & Figuereo Benítez, J. C. (2024). El Big Data y la inteligencia artificial como soluciones a la desinformación. Doxa Comunicación, 38, 437–451. https://doi.org/10.31921/doxacom.n38a2029.

Morosoli, S., Resendez, V., Naudts, L., Helberger, N., & de Vreese, C. (2025). "I Resist." A study of individual attitudes towards generative AI in journalism and acts of resistance, risk perceptions, trust, and credibility. Digital Journalism, 1–20. https://doi.org/10.1080/21670811.2024.2435579.

Murcia Verdú, F. J., & Lara Ramos, D. (2024). ¿Están preparadas las redacciones para la integración de la inteligencia artificial? In F. J. Murcia Verdú, & R. Ramos Antón (Coords.), La inteligencia artificial y la transformación del periodismo: Narrativas, aplicaciones y herramientas (pp. 67–89). Comunicación Social Ediciones y Publicaciones.

Møller, L. A., Skovsgaard, M., & de Vreese, C. (2024). Reinforce, readjust, reclaim: How artificial intelligence impacts journalism's professional claim. Journalism, 26(7), 1373–1390. https://doi.org/10.1177/14648849241269300.

Newman, N., Fletcher, R., Robertson, C. T., Arguedas, A. R., & Nielsen, R. K. (2024). Digital news report 2024: Key findings and analysis. Reuters Institute. Available online: https://coilink.org/20.500.12592/kprrc2m (accessed on 29 July 2025).

Nguyen, D. (2023). How news media frame data risks in their coverage of big data and AI. Internet Policy Review, 12(2), 1–30. https://doi.org/10.14763/2023.2.1708.

Noain-Sánchez, A. (2022). Addressing the impact of artificial intelligence on journalism: The perception of experts, journalists and academics. Communication & Society, 35(3), 3. https://doi.org/10.15581/003.35.3.105-121.

Oh, S., & Jung, J. (2025). Harmonizing traditional journalistic values with emerging AI technologies: A systematic review of journalists' perception. Media and Communication, 13, 1-27. https://doi.org/10.17645/mac.9495.

Palomo, B., Tandoc. E.C., Jr., & Cunha, R. (2023). El impacto de la desinformación en las rutinas profesionales y soluciones basadas en la inteligencia artificial. Estudios sobre el Mensaje Periodístico 29(4), 757–759. https://dx.doi.org/10.5209/esmp.92160.

Papadimitriou, A. (2016). The future of communication: Artificial intelligence and social networks. Media & communication studies. Mälmo University. Available online: http://www.diva-portal.org/smash/record.jsf?pid=diva2%3A1481794&dswid=5239 (accessed on).

Peña Fernández, S., Meso Ayerdi, K., Larrondo Ureta, A., & Díaz Noci, J. (2023). Without journalists, there is no journalism: The social dimension of generative artificial intelligence in the media. Profesional de la Información, 32(2), e320227. https://doi.org/10.3145/epi.2023.mar.27.

Peña-Fernández, S., Peña-Alonso, U., & Eizmendi-Iraola, M. (2023). El discurso de los periodistas sobre el impacto de la inteligencia artificial generativa en la desinformación. Estudios sobre el Mensaje Periodístico, 29(4), 833–841. https://doi.org/10.5209/esmp.88673.

Pihlajarinne, T., & Alén-Savikko, A. (2022). Artificial intelligence and the media: Reconsidering rights and responsibilities. Edward Elgar Publishing. Available online: https://tinyurl.com/2zdbrepu (accessed on).

Pilo García, M. A., Romero Gutiérrez, J. M., de Casas Moreno, P., & Aguaded, I. (2024). El impacto de la Inteligencia Artificial en Comunicación. Revisión sistematizada de la producción científica española en Scopus (2020–2023). Razón y Palabra, 28(119), 65–79. https://doi.org/10.26807/rp.v28i119.2098.

Pérez, F., Broseta, B., Escribá, A., López, G., Maudos, J., & Pascual, F. (2023). Los medios de comunicación en la era digital. Fundación BBVA.

Rick, J., & Hanitzsch, T. (2023). Journalists' perceptions of precarity: Toward a theoretical model. Journalism Studies, 25(2), 199–217. https://doi.org/10.1080/1461670X.2023.2293827.

Rodríguez Martelo, T., Rúas Araújo, J., & Maroto González, I. (2023). Innovation, digitization, and disinformation management in European regional television stations in the Circom network. Profesional de la Información, 32(1), 1-14. https://doi.org/10.3145/epi.2023.ene.12.

Ross Arguedas, A. (2024). Public attitudes towards the use of AI in journalism. Reuters Institute for the Study of Journalism, University of Oxford. Available online: https://reutersinstitute.politics.ox.ac.uk/digital-news-report/2024/public-attitudes-towards-use-ai-and-journalism (accessed on).

Rostamian, S. & Moradi Kamreh, M. (2024). AI in Broadcast Media Management: Opportunities and Challenges. AI and Tech in Behavioral and Social Sciences, 2(3), 21-28. https://doi.org/10.61838/kman.aitech.2.3.3

Rubin, V. L. (2022). Misinformation and disinformation: Detecting fakes with the eye and AI. Springer.

Saeidnia, H. R., Hosseini, E., Lund, B., Tehrani, M. A., Zaker, S., & Molaei, S. (2025). Artificial intelligence in the battle against disinformation and misinformation: A systematic review of challenges and approaches. Knowledge and Information Systems, 67(4), 3139–3158. https://doi.org/10.1007/s10115-024-02337-7.

Sanguinetti, P. A. B. L. O. (2023). Inteligencia artificial en periodismo: Oportunidades, riesgos, incógnitas. Cuadernos de periodistas: Revista de la Asociación de la Prensa de Madrid, 46, 9–17.

Santos, F. C. C. (2023). Artificial intelligence in automated detection of disinformation: A thematic analysis. Journalism and Media, 4(2), 679–687. https://doi.org/10.3390/journalmedia4020043.

Sarrionandia, B., Peña-Fernández, S., Ángel Pérez-Dasilva, J., & Larrondo-Ureta, A. (2025). Artificial intelligence training in media: Addressing technical and ethical challenges for journalists and media professionals. Frontiers in Communication, 10, 1537918. https://doi.org/10.3389/fcomm.2025.1537918.

Schaetz, N., Schjøtt, A., Dodds, T., & Mellado, C. (2025). The (invisible) work involved in bridging the research-practice gap in journalism. Journalism, 26(8), 1591-1602. https://doi-org.ehu.idm.oclc.org/10.1177/14648849251343964

Sixto García, J., Rodríguez Vázquez, A. I., & López García, X. (2021). Sistemas de verificación en medios nativos digitales e implicación de la audiencia en la lucha contra la desinformación en el modelo ibérico. Revista de comunicación de la SEECI, 54, 41–61. https://doi.org/10.15198/seeci.2021.54.e738.

Sonni, A. F. (2025). Digital transformation in journalism: Mini review on the impact of AI on journalistic practices. Frontiers in Communication, 10, 1535156. https://doi.org/10.3389/fcomm.2025.1535156.

Sonni, A. F., Hafied, H., Irwanto, I., & Latuheru, R. (2024). Digital newsroom transformation: A systematic review of the impact of artificial intelligence on journalistic practices, news narratives, and ethical challenges. Journalism and Media, 5(4), 1554–1570. https://doi.org/10.3390/journalmedia5040097.

Subiela Hernández, B. H., Gómez Company, A., & Vizcaíno Laorga, R. (2023). Retos y oportunidades en la lucha contra la desinformación y los derechos de autor en Periodismo: MediaVerse (IA, blockchain y smart contracts). Estudios sobre el Mensaje Periodístico, 29(4), 869–880. https://doi.org/10.5209/esmp.88081.

Tejedor, S., & Sancho Ligorred, B. (2023). Cartografía mundial de herramientas, fact-checkers y proyectos contra la infodemia. Estudios sobre el Mensaje Periodístico, 29(4), 933–942. https://doi.org/10.5209/esmp.92160.

Thurman, N., Dörr, K., & Kunert, J. (2017). When reporters get hands-on with robo-writing: Professionals consider automated journalism's capabilities and consequences. Digital Journalism, 5(10), 1240–1259. https://doi.org/10.1080/21670811.2017.1289819.

Thäsler-Kordonouri, S., & Barling, K. (2023). Automated journalism in UK local newsrooms: Attitudes, integration, impact. Journalism Practice, 19(1), 58–75. https://doi.org/10.1080/17512786.2023.2184413.

Túñez López, J. M., Fieiras Ceide, C., & Vaz Álvarez, M. (2021). Impacto de la Inteligencia Artificial en el Periodismo: Transformaciones en la empresa, los productos, los contenidos y el perfil profesional. Communication & Society, 34(1), 177–193. https://doi.org/10.15581/003.34.1.177-193.

Ufarte Ruiz, M. J., Murcia Verdú, F. J., & Túñez López, J. M. (2023). Use of artificial intelligence in synthetic media: First newsrooms without journalists. Profesional de la información, 32(2), e320203. https://doi.org/10.3145/epi.2023.mar.03.

van Dalen, A. (2024). Revisiting the algorithms behind the headlines: How journalists respond to professional competition of generative AI. Journalism Practice, 1–18. https://doi.org/10.1080/17512786.2024.2389209.

Verma, D. (2024). Impact of artificial intelligence on journalism: A comprehensive review of AI in journalism. Journal of Communication and Management, 3(2), 150–156. https://doi.org/10.58966/JCM20243212.

Vrabič Dežman, D. (2024). Promising the future, encoding the past: AI hype and public media imagery. AI and Ethics, 4(3), 743–756. https://doi.org/10.1007/s43681-024-00474-x.

Wei, K., Ezell, C., Gabrieli, N., & Deshpande, C. (2024). How do AI companies "fine-tune" policy? Examining regulatory capture in AI governance. Proceedings of the AAAI/ACM Conference on AI, Ethics, and Society, 7(1), 1539–1555. https://doi.org/10.1609/aies.v7i1.31745.

Wilczek, B., Haim, M., & Thurman, N. (2024). Transforming the value chain of local journalism with artificial intelligence. AI Magazine, 45(2), 200–211. https://doi.org/10.1002/aaai.12174.

Wu, S. (2024). Journalists as individual users of artificial intelligence: Examining journalists' "value-motivated use" of ChatGPT and other AI tools within and without the newsroom. Journalism. https://doi.org/10.1177/14648849241303047.

Wu, S., Tandoc, E. C., & Salmon, C. T. (2019). Journalism reconfigured. Journalism Studies, 20(10), 1440–1457. https://doi.org/10.1080/1461670X.2018.1521299.

Zheng, Y., Zhong, B., & Yang, F. (2018). When algorithms meet journalism: The user perception to automated news in a cross-cultural context. Computers in Human Behavior, 86, 266–275. https://doi.org/10.1016/j.chb.2018.04.046.